\documentclass[aps,prc,12pt,nofootinbib,superscriptaddress,showkeys,english]{revtex4-2}
\usepackage[english]{babel}
\usepackage{graphicx}
\usepackage{color}
\usepackage{amsmath}
\usepackage{slashed}
\usepackage{amssymb}
\usepackage{multirow}
\usepackage{color}
\usepackage{appendix}
\usepackage{enumitem}

\begin{document}
\title{Search for dark photons in heavy-ion collisions}
\author{A. W. Romero Jorge}
\email{aromerojorge@physik.uni-frankfurt.de}
\affiliation{Frankfurt Institute for Advanced Studies, Ruth Moufang Str. 1, 60438 Frankfurt, Germany}
		
\affiliation{Institute for Theoretical Physics, Johann Wolfgang Goethe University, Max-von-Laue-Str. 1, 60438 Frankfurt am Main, Germany }
	
\affiliation{Helmholtz Research Academy Hessen for FAIR (HFHF), GSI Helmholtz	Center for Heavy Ion Physics. Campus Frankfurt, 60438 Frankfurt, Germany}
\affiliation{Instituto de Cibern\'{e}tica, Matem\'{a}tica y F\'{\i}sica (ICIMAF), Calle E esq a 15 Vedado 10400 La Habana Cuba}
	\author{Elena Bratkovskaya}
\affiliation{GSI Helmholtzzentrum für Schwerionenforschung GmbH, Planckstr. 1, 64291 Darmstadt, Germany}
\affiliation{Institute for Theoretical Physics, Johann Wolfgang Goethe University, Max-von-Laue-Str. 1, 60438 Frankfurt am Main, Germany }
\affiliation{Helmholtz Research Academy Hessen for FAIR (HFHF), GSI Helmholtz	Center for Heavy Ion Physics. Campus Frankfurt, 60438 Frankfurt, Germany}
\author{Laura Sagunski}
\affiliation{Institute for Theoretical Physics, Johann Wolfgang Goethe University, Max-von-Laue-Str. 1, 60438 Frankfurt am Main, Germany }

\date{\today}

\begin{abstract}
The vector $U$-bosons, or so called 'dark photons', are one of the possible candidates for the dark matter mediators. We present a procedure to derive theoretical constraints on the upper limit of kinetic mixing parameter $\epsilon^2(M_U)$ from heavy-ion as well as $p+p$ and $p+A$ dilepton data from SIS to LHC energies. Our study is based on the microscopic Parton-Hadron-String Dynamics (PHSD) transport approach which reproduces  the measured dilepton spectra in $p+p$, $p+A$ and $A+A$ collisions well. In addition to the different dilepton channels originating from interactions and decays of ordinary Standard Model matter particles (mesons and baryons), we incorporate in the PHSD the decay of hypothetical $U$-bosons to dileptons, $U \to e^+ e^-$, where the $U$-bosons themselves are produced by the Dalitz decay of pions, $\eta$-mesons, Delta resonances as well as by vector meson and $K^+$ decays. This analysis provides the upper limit on the $\epsilon^2(M_U)$ and  can also help to estimate the requested accuracy for future experimental searches of 'light' dark photons by dilepton experiments.
\end{abstract}
\maketitle

\section{Introduction}

The search for dark matter (DM) candidates in Earth experiments, using the possible coupling of dark matter particles and Standard Model (SM) particles, is one of the exciting and challenging directions of modern physics.
The vector portal involves a mixing between the $U(1)$ and $U(1)^\prime$ gauge symmetries,
 where the Lagrangian is characterized by an interaction between the field-strength tensors of the Standard Model photon and the dark matter  vector bosons: ${\cal L} \sim \epsilon^2/2 , F_{\mu\nu}{F^{\mu\nu}}^\prime$. The mediators in this case are vector $U$-bosons, also referred to as 'dark photons' or 'hidden photons', with an unknown mass $M_U$. The kinetic mixing parameter $\epsilon^2$ governs the strength of the interactions between dark matter and Standard Model particles.
This mixing allows $U$-bosons to decay into lepton pairs, such as $e^+e^-$ or $\mu^+\mu^-$. Light $U$-bosons can be generated through the decays of Standard Model particles, particularly Dalitz decays of pseudoscalar mesons ($\pi^0$, $\eta$) and baryonic resonances like the $\Delta$. These decays present potential avenues to detect dark photons in dilepton experiments, sparking significant experimental and theoretical interest \cite{Alexander:2016aln,Battaglieri:2017aum}.

This study builds upon previous work (cf. \cite{Schmidt:2021hhs}) that investigated the upper limit of the kinetic mixing parameter $\epsilon^2(M_U)$ for hypothetical $U$-bosons with a mass $M_U \leq 0.6$ GeV, extending the analysis to cover a mass range up to 2 GeV. 
Beyond the previously examined Dalitz decays of $\pi^0$, $\eta$, and $\Delta$ particles, this study introduces additional dark photon production channels, including the Dalitz decay of the $\omega$, direct decays of vector mesons such as the $\rho$, $\omega$, and $\phi$, as well as decays from $K^+$ mesons.

\section{Standard matter production in the PHSD}

Parton-Hadron-String Dynamics (PHSD) is a microscopic, non-equilibrium transport model that integrates both hadronic and partonic degrees-of-freedom \cite{Cassing:2008nn,Moreau:2019vhw}. 
It provides a comprehensive description of relativistic heavy-ion collisions, covering the entire process from the initial out-of-equilibrium nucleon-nucleon collisions, through the formation and interactions of the quark-gluon plasma (QGP), and extending to the hadronization and final-state interactions of the produced hadrons. The time evolution of the system is governed by the Cassing-Juchem generalized off-shell transport equations in a test-particle framework, based on the Kadanoff-Baym equations in first-order gradient expansion, which are suitable for describing strongly interacting degrees of freedom \cite{Cassing:2008nn}.
The PHSD model has been effectively used to study $p+p$, $p+A$, and $A+A$ collisions from SIS18 to LHC energies, accurately reproducing a large range of hadronic observables, including dileptons and photons \cite{Cassing:2008nn,Moreau:2019vhw}.

\subsection{Dilepton production from the SM sources in the PHSD}

Dileptons, such as $e^+e^-$ and $\mu^+\mu^-$ pairs, can be emitted from virtual photon decays throughout all stages of heavy-ion collisions, originating from hadronic and partonic sources. The hadronic sources include Dalitz decays of mesons and baryons (e.g., $\pi^0$, $\eta$, $\Delta$) and direct decays of vector mesons (e.g., $\rho$, $\omega$, $\phi$).
The dileptons from correlated $D-$meson decays and 'thermal' radiation from the QGP
are included also - see Ref. \cite{Bratkovskaya:2013vx,Song:2018xca} for details of the dilepton evaluation.

\section{Dark photon production and dilepton decay channels  in the PHSD}

In this study, we follow the strategy used in our previous work \cite{Schmidt:2021hhs}. Alongside the previously considered Dalitz decays of $\pi^0 \to \gamma + U$, $\eta \to \gamma + U$, and $\Delta \to N + U$, we incorporate additional channels for dark photon production. These include direct decays of vector mesons $V \to U$, where $V = \rho, \phi, \omega$, as well as the $\omega$ Dalitz decay $\omega \to \pi^0 + U$ and the kaon decay $K^+ \to \pi^+ + U$.

The dilepton yield from a $U$-boson decay of mass $M_U$ can be evaluated as the sum of all possible contributions 
for a given mass $M_U$:
\begin{eqnarray}
N^{U\to e^+e^-} = 
\sum_{h} N^{U\to e^+e^-}_{h}  
 = Br^{U\to e^+e^-} 
 \sum_{h}
 N_{h \to  U X},
\end{eqnarray}\label{NUee}
where $Br^{U\to e^+e^-}$ is the branching ratio for the decay of $U$-bosons to $e^+e^-$ and, $X = \gamma$ for $h = \pi, \eta, \omega$, $X = N$ for $h = \Delta$, and $X = \pi$ for $h = \omega, K^+$. 
We assume that the width of the $U$-boson is zero (or very small), i.e., it contributes only 
to a single $dM$ bin of dilepton spectra from SM sources. 
On the other hand,  the yield of $U$-bosons  of mass $M_U$ themselves can be
estimated from the coupling to the virtual photons \cite{HADES:2013nab}:
\begin{eqnarray}
\footnotesize{
 N_{h\to UX} = N_h Br_{h\to \gamma X}  \cdot    
   \frac{\Gamma_{h\to UX}}{\Gamma_{h\to \gamma X}}, }\label{mNU2}
\end{eqnarray}
where $X=\pi,\eta$ for $h=\gamma$ and $X=\Delta$ for $h=N$. The ratio of the partial widths for the Dalitz decays of $\pi^0$ and $\eta$ mesons into $U$-bosons and real photons can be evaluated, as taken from Refs. \cite{Batell:2009yf,Batell:2009di}.
For the evaluation of the partial decay widths of a broad $\Delta$  resonance, one has to take into account the $\Delta$ spectral function $A(M_\Delta)$ as used also in the HADES study \cite{HADES:2013nab}. 
For other dark photon decays with $h=V$, $\omega$, and $K^+$, the partial widths are taken from the models developed in Refs. \cite{Batell:2009di}, \cite{Gorbunov:2024nph}, and \cite{Pospelov:2008zw}, respectively.

The branching ratio for the decay of $U$-bosons to $e^+e^-$, entering Eq. (\ref{NUee}), is 
adopted from Ref. \cite{Batell:2009yf} and used also in Ref. \cite{HADES:2013nab}:
\begin{equation}
\footnotesize
   Br^{U\to ee} = \frac{\Gamma_{U \rightarrow e^+e^-}}{\Gamma_{tot}^U} \label{Bree}
    = \left(1 + \sqrt{1 - \frac{4m_\mu^2}{M_U^2}} \left( 1 + \frac{2m_\mu^2}{M_U}\right) \left(1 + R(M_U)\right)\right)^{-1},  
\end{equation}
where $m_\mu$ is the muon mass. The total decay width of a $U$-boson is the sum of the partial decay widths 
to hadrons, $e^+e^-$ and $\mu^+\mu^-$ pairs,
$\Gamma_{tot}^U= \Gamma_{U\to hadr} + \Gamma_{U\to e^+e^-} + \Gamma_{U\to\mu^+\mu^-}$.
The expression (\ref{Bree}) has been evaluated using that $\Gamma_{U\to\mu^+\mu^-} = \Gamma_{U\to e^+e^-}$ 
due to lepton universality for $M_U\gg 2m_\mu$. The hadronic decay width of $U$-bosons 
is chosen such that $\Gamma_{U\to hadr} = R(\sqrt{s}=M_U)\Gamma_{U\to\mu^+\mu^-}$, where 
the factor $R(\sqrt{s}) = \sigma_{e^+e^-\rightarrow hadrons}$/$\sigma_{e^+e^-\rightarrow \mu^+\mu^-}$.
Eq. (\ref{Bree}) is only valid for $M_U<0.6$ GeV, and an extension of the branching ratio $Br^{U\to ee}$ up to $M_U<2$ GeV is taken from \cite{Liu:2014cma}.

\section{Results for the dilepton spectra from U-boson decays and constraints on $\epsilon^2 (M_U)$}

Given that both the kinetic mixing parameter $\epsilon^2$ and the mass of the $U$-boson are unknown, we adopt the following method to place constraints on $\epsilon^2(M_U)$. For each dilepton mass bin $dM$, which is chosen as 10 MeV in our simulations, we compute the integrated dilepton yield from $U$-bosons with masses within the range $[M_U, M_U + dM]$ using Eq. (\ref{NUee}) and normalize it by the bin width $dM$. The resulting dilepton yield per bin $dM$ is denoted as $dN^{sumU}/dM$, representing the sum of all kinematically accessible dilepton contributions from $U$-bosons produced via $h\to \gamma X$ dark photon production channels. 
Assuming that $\epsilon^2$ remains constant across $dM$, we express the dilepton yield as 
$dN^{sumU}/dM = \epsilon^2 \cdot dN^{sumU}_{\epsilon=1}/dM$, where $dN^{sumU}_{\epsilon=1}/dM$ represents the dilepton yield computed with $\epsilon = 1$.
The total yield from all possible dilepton sources, including both SM channels and $U$-boson decays, can then be written as:
\begin{equation}
\footnotesize
    \frac{dN}{dM}^{total} = \frac{dN}{dM}^{sum SM} + \frac{dN}{dM}^{sum U} = \frac{dN}{dM}^{sum SM} + \epsilon^2\frac{dN_{\epsilon=1}^{sum U}}{dM}.
\end{equation}\label{dNdMepsil}  

We can now impose constraints on $\epsilon^2(M_U)$ by requiring that the total yield $dN^{total}/dM$ does not exceed the SM contribution by more than a certain fraction $C_U$ in each bin $dM$. Here, $C_U$ sets the maximum additional dilepton yield from dark photons allowed over the SM yield (e.g., $C_U = 0.1$ implies that dark photons contribute an additional 10\% to the SM yield). This condition is expressed as:
$  \frac{dN}{dM}^{total} = (1+C_U) \frac{dN}{dM}^{sum SM}$.
Combining the last equation and  (\ref{dNdMepsil}), one obtains that 
the kinetic mixing parameter $\epsilon^2$ for $M_U$ can be evaluated as 
\begin{equation}
\footnotesize
    \epsilon^2 (M_U) = C_U \cdot  \left. { \left(\frac{dN}{dM}^{sumSM} \right)} \right/
    {\left(\frac{dN_{\epsilon=1}^{sum U}}{dM} \right)}.
\end{equation}\label{epsM}

Eq. (\ref{epsM}) allows the determination of $\epsilon^2$ for each mass bin $[M_U, M_U + dM]$, representing the weighted dilepton yield from dark photons relative to SM channels. The procedure also takes into account the experimental acceptance for $e^+e^-$ pairs from $U$-boson decays in the same way as for SM channels, making it suitable for comparison with experimental results. By this comparison we can explore the possible range of $C_U$, which controls the extra contribution from dark photons. Given that dark photons have not been observed in dilepton experiments, this additional yield must remain within the experimental uncertainties, assuming that the SM predictions agree with the data.

\begin{figure}
\includegraphics[width=0.49\textwidth]{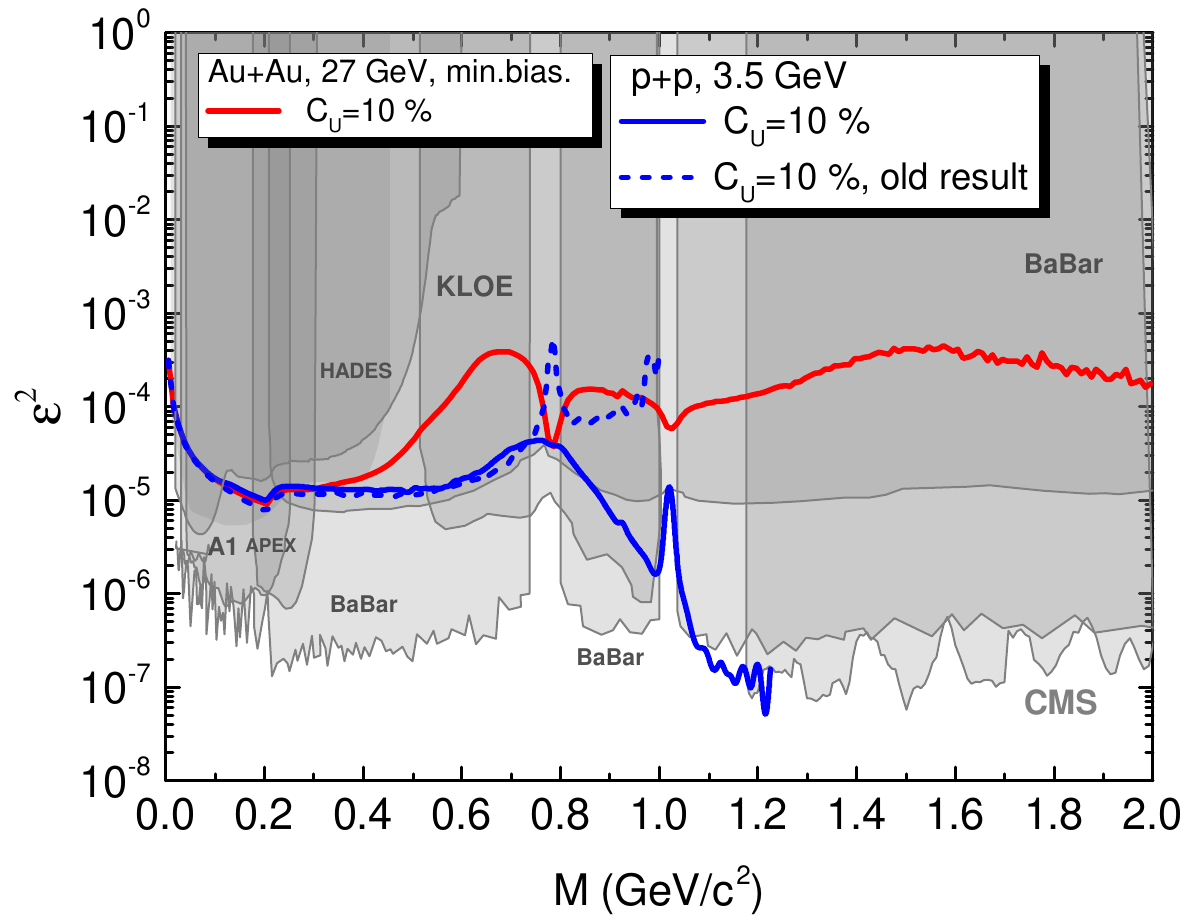}
\includegraphics[width=0.49\textwidth]{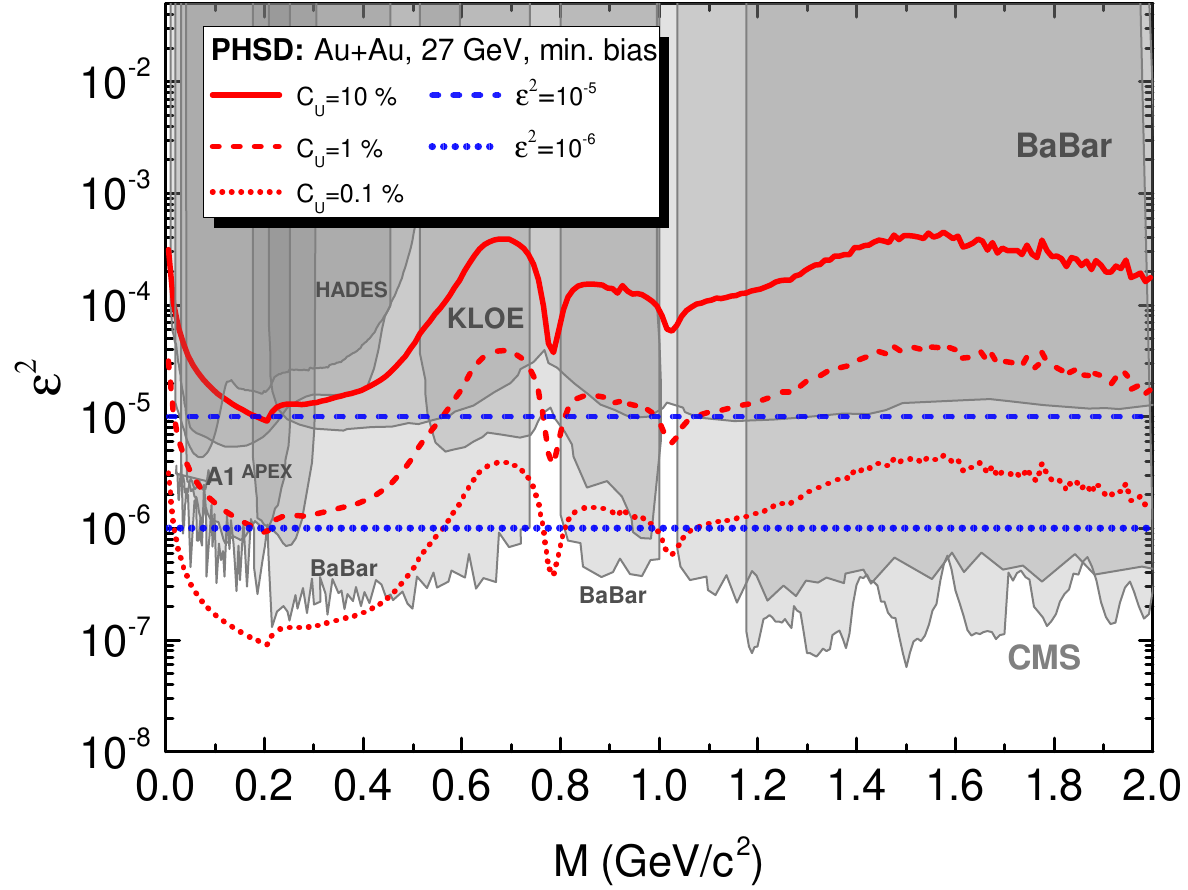}
\caption{
\textbf{Left}: The kinetic mixing parameter $\epsilon^2$ extracted from the PHSD dilepton spectra for $p+p$ at 3.5 GeV (blue line), $Au+Au$ at 27 GeV (red line) in a comparison to  the compilation of the worldwide data \cite{Battaglieri:2017aum}.
The preliminary PHSD results are shown with 10\% allowed surplus of the $U$-boson contributions over the total SM. We include our previous result for $p + p$ collisions at 3.5 GeV (dashed blue line) for comparison with the current results.
\textbf{Right}: The kinetic mixing parameter $\epsilon^2$ extracted from the PHSD dilepton spectra for $Au+Au$ at 27 GeV calculated for different $\epsilon^2$ scenarios (red).
The PHSD results are shown with 0.1, 1, 10\% allowed surplus of the $U$-boson contributions over 
the total SM yield (cf. color coding in the legend). The dotted and dashed blue line shows, the constant $\epsilon^2 =10^{-6}$ and $\epsilon^2 =10^{-5}$, respectively. 
}
\label{epsil2}
\end{figure}

In the left panel of Fig. \ref{epsil2}, we show the (preliminary) results for the kinetic mixing parameter $\epsilon^2$ versus $M_U$ extracted from the PHSD dilepton spectra for $p+p$ at 3.5 GeV (blue line), 
$Au+Au$ at 27 GeV (red line) in comparison with the world data compilation (gray background) \cite{Battaglieri:2017aum}.
The PHSD results are shown with a 10\% allowed surplus of the $U$-boson contributions over the total SM yield.
We present our calculations for larger $M_U$ values to illustrate that our theoretical approach can yield valuable constraints on $\epsilon^2$ across a range of $M_U$.
It is important to note that the shape of the theoretically derived $\epsilon^2(M_U)$ curve remains unaffected by the experimental detector acceptance, as this factor influences both SM and DM contributions equally at a given mass $M = M_U$. For smaller $M_U$ values (i.e., $M_U < m_{\pi^0}$), the extracted $\epsilon^2$ shows minimal dependence on the collision system size and energy, primarily because the Dalitz decay of $\pi^0$ is the dominant decay channel. However, as $M_U$ increases, additional decay channels become accessible, and the resulting constraints on $\epsilon^2$ are influenced by the proportion of $U$-boson production channels relative to all other dilepton production channels. We found that the extracted upper limit of $\epsilon^2(M_U)$ for $p+p$ at 3.5 GeV is consistent with the experimental results of the BaBar, KLOE experiment between 0.2 < $m_U $ < 1 GeV with $C_U=10\%$.
 
In the right panel of Fig. \ref{epsil2}  shows  the kinetic mixing parameter $\epsilon^2$ 
extracted from the PHSD dilepton spectra for $Au+Au$ at 27 GeV. 
The PHSD results are displayed with 0.1, 1, 10\% allowed surplus of the $U$-boson contributions over  the total SM yield.  The results calculated with $C_U = 1\%$ are close to the A1
experimental result for $0.01 < M_U < 0.3$ GeV.  The dotted and dashed blue line shows the constant $\epsilon^2 =10^{-6}$ and $\epsilon^2 =10^{-5}$ respectively, which approximately corresponds to the present knowledge on the upper limit based on the compilation of the worldwide experiments \cite{Battaglieri:2017aum}.

\section{Summary}
This study presents a microscopic transport calculation using the PHSD approach to estimate the dilepton yield from hypothetical dark photons (or $U$-bosons) decaying to $e^+e^-$ in $p+p$, $A+A$ collisions at relativistic energies. It extends the previous investigation \cite{Schmidt:2021hhs} - which included the dark photon production   from $\pi^0 \to \gamma + U$, $\eta \to \gamma + U$, and $\Delta \to N + U$ -
by accounting for the additional channels such as a direct decay of vector mesons $V \to U$, where $V = \rho, \phi, \omega$, as well as the $\omega$ Dalitz decay, $\omega \to \pi^0 + U$, and the kaon decay, $K^+ \to \pi^+ + U$.
In order to define the theoretical constraints on the upper limit of the kinetic mixing parameter $\epsilon^2(M_U)$, we employ the procedure introduced  in Ref. \cite{Schmidt:2021hhs}: since dark photons are not observed in dilepton experiments so far, we can require that their contribution can not exceed some limit (defined by "surplus" $C_U$), which would make them visible in the dilepton experimental data.

Our analysis reveals that the upper limit on $\epsilon(M_U)$ is in agreement with the experimental results from the BaBar and KLOE experiments for the mass range $0.2 < M_U < 1$ GeV with a surplus factor $C_U = 10\%$, as well as with the A1 experiment for $0.01 < M_U < 0.3$ GeV with $C_U = 1\%$. This result also matches the global compilation of experimental data.

\vspace*{1mm}
\section*{Acknowledgements}

 A.R.J. acknowledges the financial support provided by the Stiftung Giersch.
We acknowledge support by the Deutsche Forschungsgemeinschaft (DFG)
through the grant CRC-TR 211 'Strong-interaction matter under extreme conditions' - Project number 315477589 - TRR 211 and the CNRS Helmholtz Dark Matter Lab (DMLab). The computational resources have been provided by the Center for Scientific Computing (CSC) of  Goethe University Frankfurt.

\end{document}